\documentclass[aps,prl,twocolumn,showpacs,superscriptaddress]{revtex4}
\usepackage{graphicx}
\usepackage{amsmath}
\usepackage{color}

\begin{document}

\title{Polaron-induced lattice distortion of (In,Ga)As quantum dots by optically excited carriers}

\author{S.\ Tiemeyer}
\email{sebastian.tiemeyer@tu-dortmund.de}
\affiliation{Fakult\"at Physik / DELTA, Technische Universit\"at Dortmund, D-44221 Dortmund, Germany}
\author{M.\ Bombeck}
\affiliation{Experimentelle Physik II, Technische Universit\"at Dortmund, D-44221 Dortmund, Germany}
\author{M.\ Paulus}
\affiliation{Fakult\"at Physik / DELTA, Technische Universit\"at Dortmund, D-44221 Dortmund, Germany}
\author{C.\ Sternemann}
\affiliation{Fakult\"at Physik / DELTA, Technische Universit\"at Dortmund, D-44221 Dortmund, Germany}
\author{J.\ Nase}
\affiliation{Fakult\"at Physik / DELTA, Technische Universit\"at Dortmund, D-44221 Dortmund, Germany}
\author{H.\ G\"ohring}
\affiliation{Fakult\"at Physik / DELTA, Technische Universit\"at Dortmund, D-44221 Dortmund, Germany}
\author{F.\,J.\ Wirkert}
\affiliation{Fakult\"at Physik / DELTA, Technische Universit\"at Dortmund, D-44221 Dortmund, Germany}
\author{J.\ M\"oller}
\affiliation{Fakult\"at Physik / DELTA, Technische Universit\"at Dortmund, D-44221 Dortmund, Germany}
\author{T.\ B\"uning}
\affiliation{Fakult\"at Physik / DELTA, Technische Universit\"at Dortmund, D-44221 Dortmund, Germany}

\author{O.\,H.\ Seeck}
\affiliation{Deutsches Elektronen-Synchrotron DESY, D-22607 Hamburg, Germany}
\author{D.\ Reuter}
\altaffiliation[Present address: ]{Optoelektronische Materialien und Bauelemente, Universit\"at Paderborn, D-33098 Paderborn, Germany}
\affiliation{Angewandte Festk\"orperphysik, Ruhr-Universit\"at Bochum, D-44780 Bochum, Germany}
\author{A.\,D.\ Wieck}
\affiliation{Angewandte Festk\"orperphysik, Ruhr-Universit\"at Bochum, D-44780 Bochum, Germany}
\author{M.\ Bayer}
\affiliation{Experimentelle Physik 2, Technische Universit\"at Dortmund, D-44221 Dortmund, Germany}
\author{M.\ Tolan}
\affiliation{Fakult\"at Physik / DELTA, Technische Universit\"at Dortmund, D-44221 Dortmund, Germany}

\date{\today}

\begin{abstract}
We report on a high resolution x-ray diffraction study unveiling the
effect of carriers optically injected into (In,Ga)As quantum dots on
the surrounding GaAs crystal matrix. We find a tetragonal lattice
expansion with enhanced elongation along the [001] crystal axis that
is superimposed on an isotropic lattice extension. The isotropic
contribution arises from excitation induced lattice heating as
confirmed by temperature dependent reference studies. The tetragonal
expansion on the femtometer scale is attributed to polaron formation
by carriers trapped in the quantum dots.
\end{abstract}

\maketitle

Self-assembled (In,Ga)As quantum dots (QDs) are crystalline
inclusions on the scale of ten nanometers that are embedded in a
GaAs matrix. Carriers residing in QDs are three-dimensionally
confined, giving rise to discrete energy levels, similar
to atoms. The excellent optical quality of such systems has allowed
studies on fundamental problems of light-matter-interaction and has paved 
also the way of QDs into applications as
light emitters, ranging from single photon sources to high-power
laser diodes\cite{bimberg,michler,wang,benson}. An important related problem with multiple facets is
the interaction of carriers with phonons, which facilitates carrier
relaxation into the ground states after non-resonant excitation \cite{steinhoff} or
mediates also coupling to the optical modes of a resonator \cite{ates}. On the
other hand, this interaction sets limitations to the coherence of
confined charge \cite{borri} or spin excitations \cite{hernandez}.

Despite of their high relevance, carrier-phonon interactions are
still not understood in full detail. Theoretically, over the years
more and more elaborated models have been developed to account for
experimental observations, for an overview see \cite{steinhoff}. The models
range from weak coupling pictures based on (modified) Fermi's golden
rule, involving single- and two-phonon emission events\cite{bockelmann,inoshita,verguftman,ferreira}, to strong coupling descriptions leading to polaron formation and involving
quantum kinetic effects \cite{inoshita2,kral,verzelen}. {Typically}, the
carrier-phonon interaction has been assessed through intra- or
interband optical transitions in QDs. \cite{hameau,kurtze} For example, non-linear
time-resolved methods like four-wave mixing revealed a drop of
coherent exciton polarization on a few ps-time scale. \cite{borri} Through the
temperature dependence of this drop, it could be uniquely related to
the interaction with acoustic phonons. In the spectral domain, this
results in spectral wings on both sides of the zero-phonon exciton
spectral line \cite{fan,borri}.

However, the impact of the coupling onto the lattice in form
of a possible distortion \cite{fan} has not been directly and quantitatively assessed so far in
experiment. For optical excitation resonant with the ground state
transition the following dynamics were predicted theoretically\cite{vagov}: A
quasi-stable polaron, which is a bound state of the injected
carriers and acoustic phonons, is formed, altering the lattice
structure and changing the lattice constant. As a result of this
distortion, a coherent phonon wave packet is emitted from the QDs,
escaping on ps-time scale into the surrounding material. Due to loss
of coherence by scattering, this wave packet eventually contributes
to heating. For non-resonant excitation, carrier relaxation towards
the ground state by phonon emission leads to additional lattice
heating\cite{hawker,bellingham}.

(In,Ga)As QDs result from heteroepitaxial, strain driven
Stranski-Krastanov growth \cite{goldstein,tersoff}. The QDs are coupled to the surrounding GaAs matrix, thereby elastically deforming the lattice. Consequently, carrier-phonon
interactions inside the QDs translate also into the GaAs crystal
lattice. Here we study the GaAs distortion due to optically excited
QD carriers by high-resolution x-ray diffraction (XRD), from which
we obtain direct, quantitative evidence for the polaron-induced
lattice expansion. While optically induced changes of bulk systems
were considered already by x-ray analysis \cite{zamponi}, this has
not been achieved so far for {nanostructures}, requiring particular
resolution and sensitivity. The XRD experiments were performed at
the beamlines BL9 \cite{krywka} and P08 \cite{seeck} of the
synchrotron radiation facilities DELTA (TU Dortmund) and PETRA III
(Deutsches Elektronen-Synchrotron DESY), respectively.

\begin{figure}[t!]
    \includegraphics[width=8.5cm]{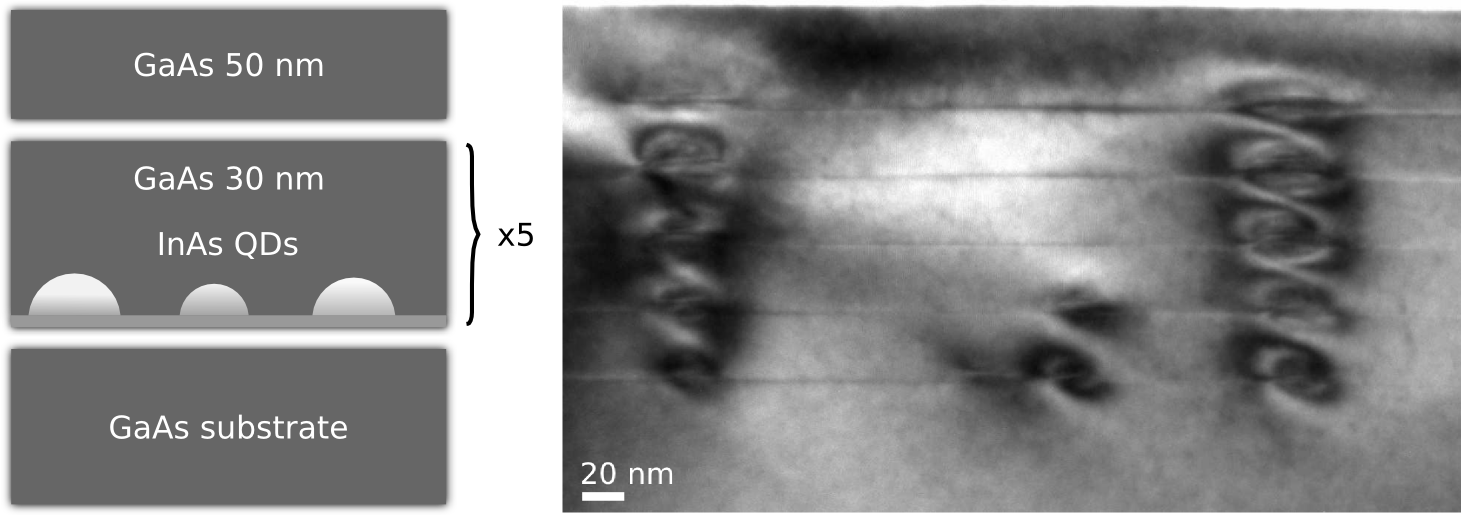}
    \vspace{2mm}
    \includegraphics[width=8.5cm]{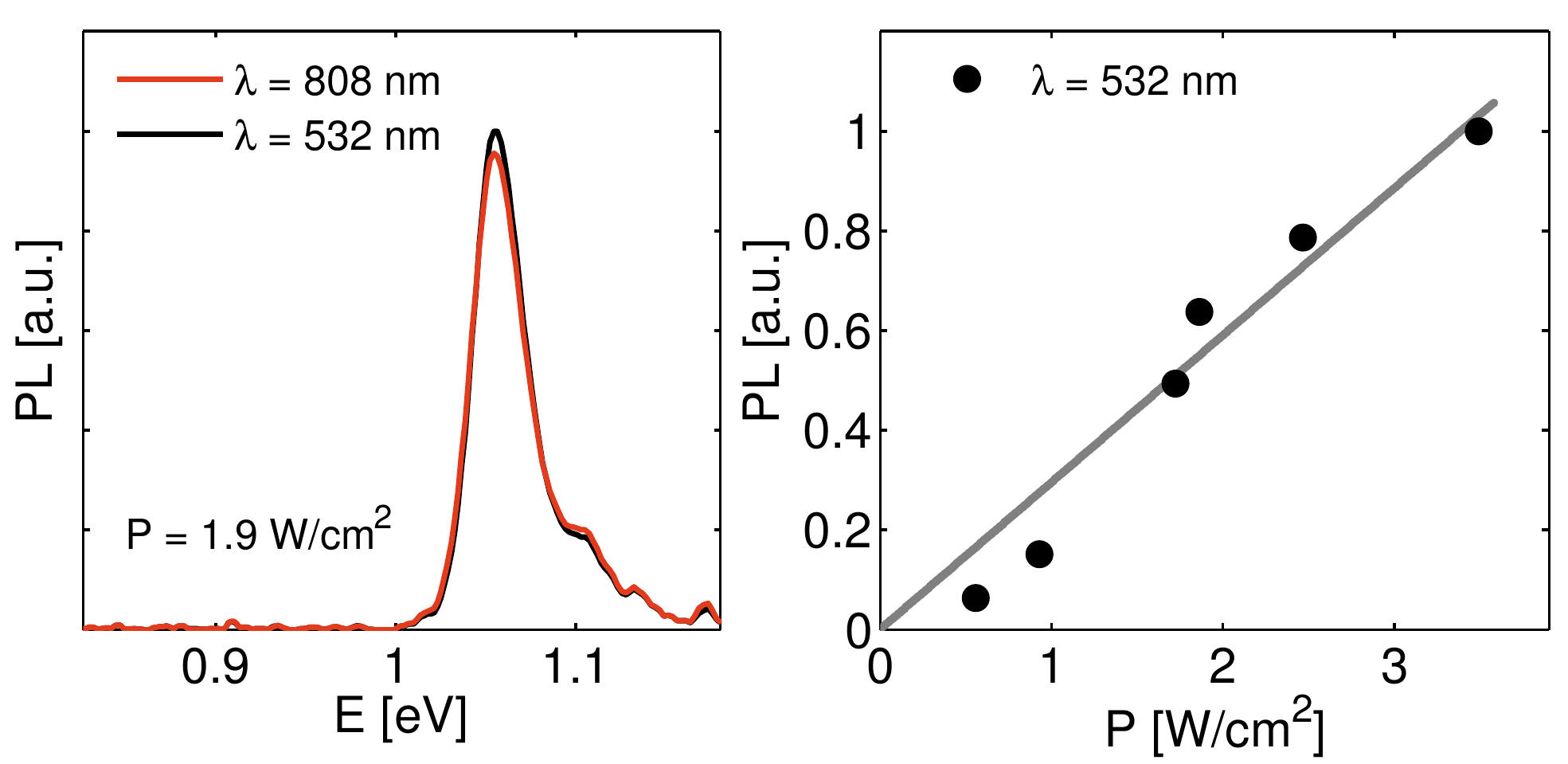}
    \caption{Sketch and X-TEM image of the QD sample (top).
    PL spectra of the QD sample using excitation laser wavelengths of 532~nm and 808~nm, respectively (bottom left). Linear dependence of PL intensity on laser power density (bottom right).}
    \label{fig:sample_setup}
\end{figure}

The study was performed on an (In,Ga)As/GaAs QD multilayer structure
grown on a (001) oriented GaAs substrate. After deposition of a 500
nm thick GaAs buffer layer, InAs corresponding to a nominal
thickness of 1.9 monolayers was deposited at a substrate temperature
of 525 $^\circ$C, resulting in the formation of the wetting layer
and the dots. The QD layer was capped by a 30~nm GaAs spacer. This
sequence was identically repeated five times. The last of these
layers was capped with 50~nm GaAs. The top part of
Figure~\ref{fig:sample_setup} shows a sketch of the sample structure
and a corresponding dark field cross-sectional transmission electron
microscope (X-TEM) image. The X-TEM image displays the stacks of
lens-shaped QDs distorting also the surrounding matrix. The dot
density is about $10^{10}$~cm$^{-2}$. The QD size is approximately
8~nm and 30~nm in height and diameter, respectively. The sample was
mounted onto the cold finger of a liquid helium continuous-flow
cryostat. The sample temperature was 100~K, large enough to avoid
carrier localization in the wetting layer, but too small to lead to
thermal carrier emission from the strongly confined ground states
(~0.4~eV total confinement potential).

\begin{figure}[t!]
    \includegraphics[width=8.5cm]{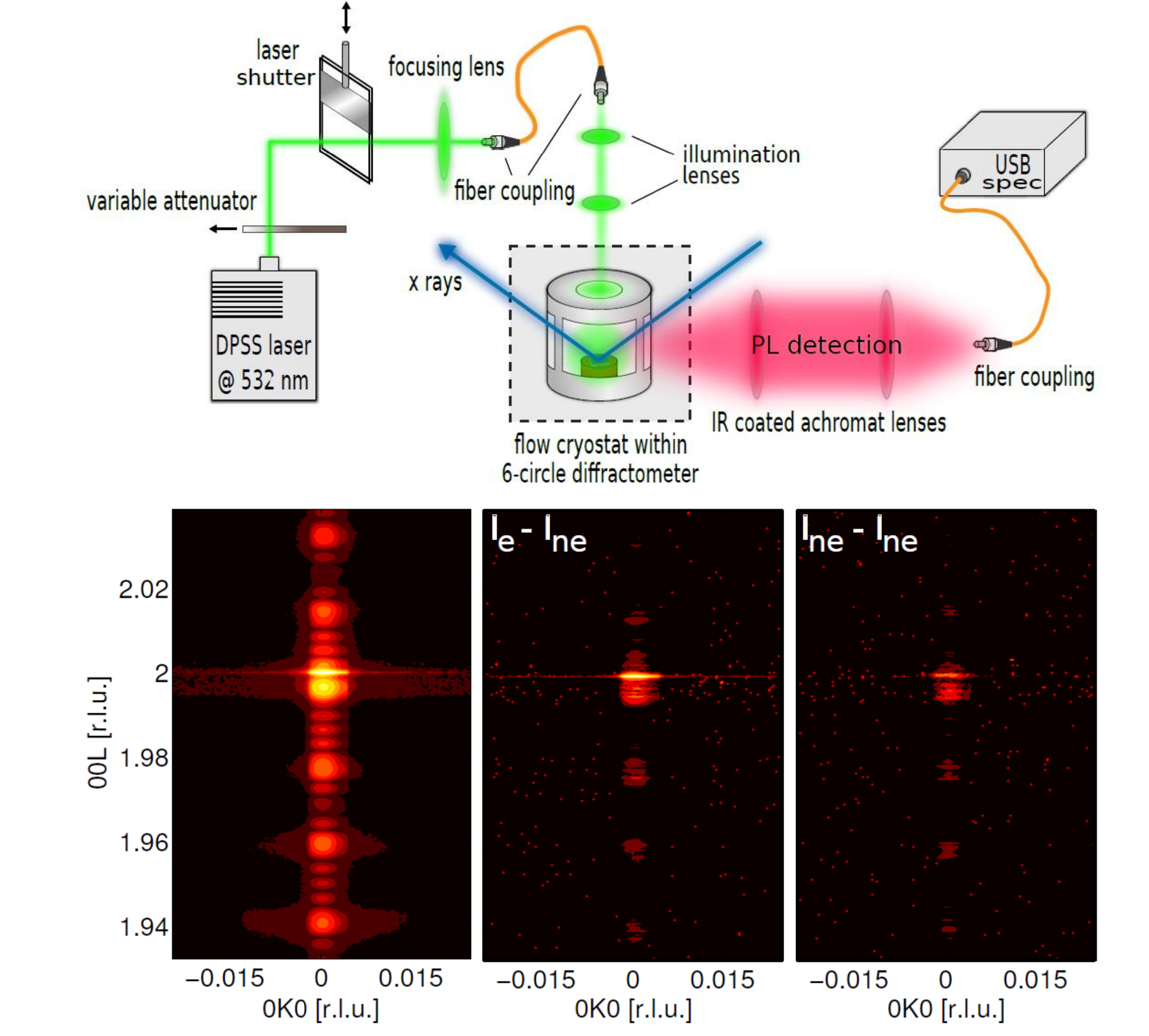}
    \caption{Top: Schematic drawing of the experimental set-up. Bottom: (left) RSM of the QD sample recorded in the vicinity of the GaAs(002) reflection. (middle) Difference of two RSMs taken for excited and non-excited sample, respectively.
(right) Reference difference of two RSMs measured each in non-excited state. }
    \label{fig:setup}
\end{figure}

The setup developed for simultaneous optical excitation and XRD
probing of the QD sample is shown in Figure~\ref{fig:setup} (top). Optical
excitation was done by a diode-pumped, continuous wave Nd:YAG laser
supplying radiation at 532~nm wavelength with 1~W maximum power. The
laser power on the sample was adjusted by a variable attenuator. The
diameter of the laser spot on the sample surface was enlarged in
order to obtain homogeneous illumination of the whole sample with
size 5 by 5 mm$^2$. Alternatively, a laser diode emitting at
$\lambda=808$~nm wavelength could be used for excitation with a
maximum power of 0.6~W. Both lasers excite electron-hole pairs into
the GaAs barriers, however, with distinctly different energies in
excess of the band gap.

Photoluminescence from the sample as result of the optical
excitation was collected by a pair of two achromatic lenses and
analyzed by a USB spectrometer. The bottom part of
Figure~\ref{fig:sample_setup} shows photoluminescence spectra
recorded with either of the two lasers at the same excitation
power while performing structural analysis, resulting in
comparable emission intensities. Thus, comparable excitation powers lead to similar carrier densities in the dots. In the applicable power range, the
intensity scales linearly with power, suggesting that the number of
excited electron-hole pairs per dot remains below unity. A laser shutter
system was added to the set-up, triggered by the beamline control.
Thereby we could record diffraction curves of the sample, both
optically excited and non-excited, within a single XRD scan
by measuring each data point twice, once with opened and once with closed laser
shutter.

The left part of Figure~\ref{fig:setup} (bottom) shows a reciprocal space map
(RSM) in the vicinity of the GaAs(002) Bragg reflection for the
non-excited QD sample, recorded at \mbox{T = 100~K}. The x-ray
photon energy was 12.38~keV. At this energy, the penetration depth of
the x-rays \cite{alsnielsen} is in the $\mu$m-range, so that effects
occuring within the InAs QD multilayer structure, located in the
top 0.2~$\mu$m layer of the sample, become accessible.

The data are scaled to reciprocal lattice units with respect to the
GaAs cubic lattice constant of $a_\text{s}=0.565$~nm at $T$~=~300 K. Along
the $L$-direction in the RSM, the GaAs(002) Bragg reflection is
accompanied by superlattice peaks originating from the layered
structure of the QD sample. An analysis of these features by
simulations utilizing the kinematical approximation connects the
superlattice peaks to the periodic wetting layer/spacing layer
system of the sample. However, as shown by the simulations, the QDs
do not contribute significantly to the x-ray scattering pattern and
can not be clearly resolved by the XRD measurements, in particular
because the dot inhomogeneities lead to a strong broadening. On the
other hand, changes by carriers trapped in the QDs affect also the
surrounding matrix, so that insight into a possible carrier-induced
lattice distortion may be taken from a GaAs reflection.

\begin{figure}[t!]
\begin{minipage}[b]{8.5cm}
    \begin{minipage}[b]{8.5cm}
    \includegraphics[width=8.5cm]{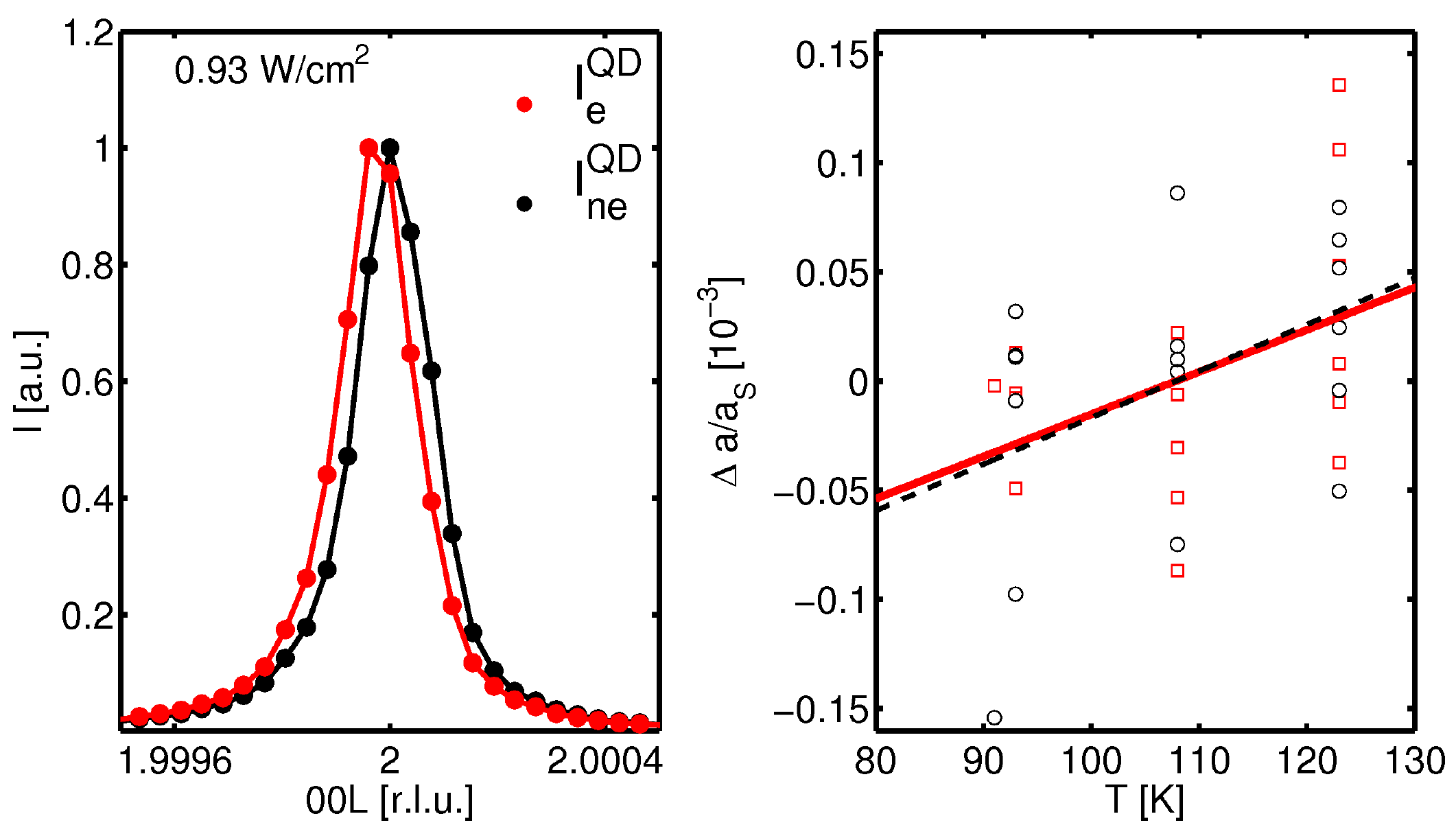}
    \includegraphics[width=9cm]{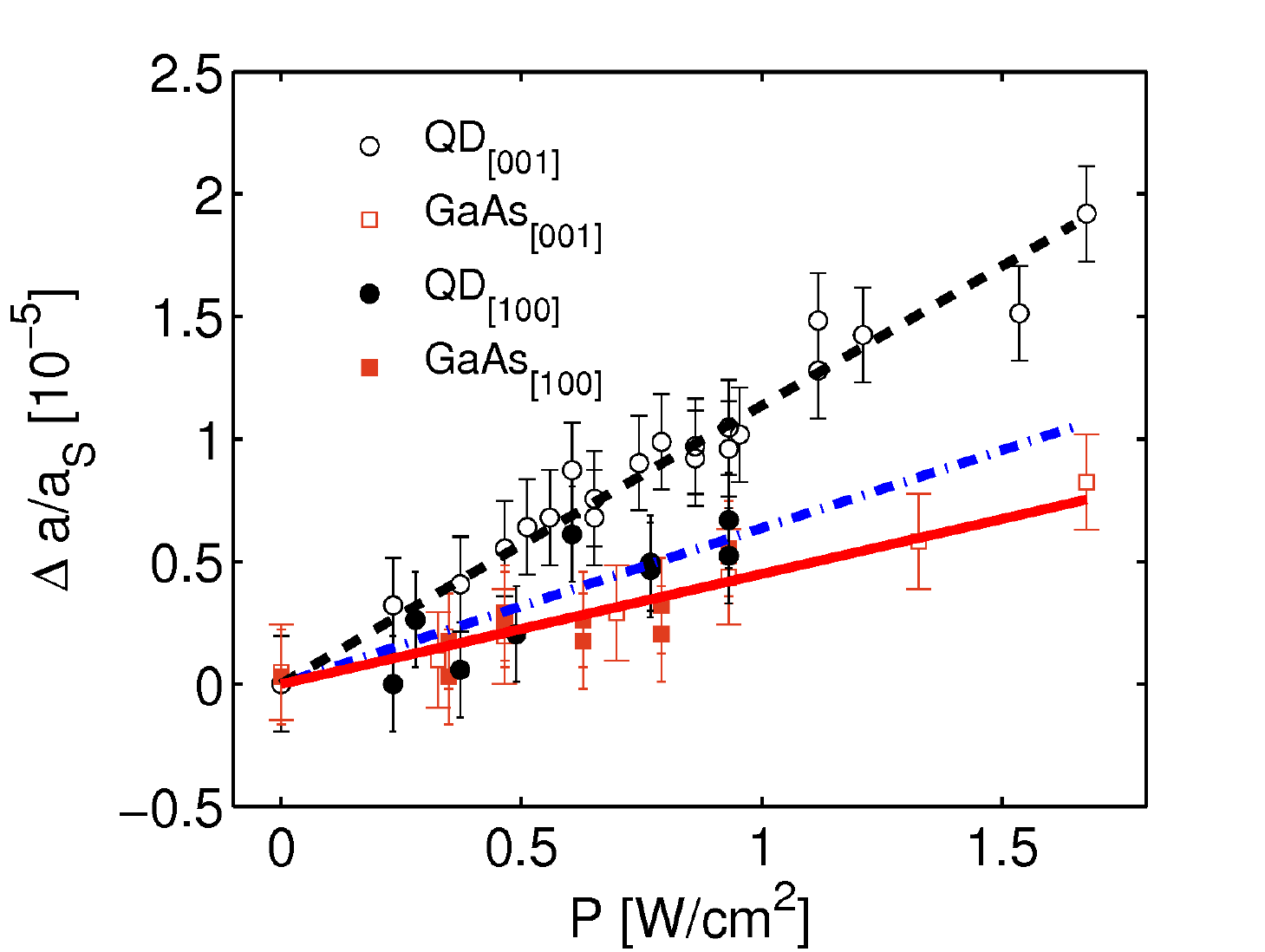}
    \caption{Peak shift of the QD (002) Bragg reflection for the excited and non-excited state at a laser power of 0.93 W/cm$^2$ (top left). Comparison of thermal expansion in [001] direction of QD (open circles) and GaAs (open squares) reference sample with the temperature varied in the range between 90~K and 125~K (top right). The solid and dashed lines correspond a linear regression to the QD and GaAs data, respectively. Dependence of relative lattice expansion along [001] and [100] crystal directions on laser excitation power density for the QD sample and GaAs reference (bottom) together with linear fits to the QD data for the [001] crystal direction (dashed), QD data in [100] direction (dashed dotted), and GaAs data in [001] and [100] directions (solid). 
   }
    \label{fig:data}
    \end{minipage}
\end{minipage}
\end{figure}

The middle part of Figure~\ref{fig:setup} gives the difference of two
RSMs, one recorded for the excited ($I_\textbf{e}$) and one for the
non-excited ($I_\textbf{ne}$) QD sample within a single scan. The
difference scattering pattern exhibits a considerable shift of the
GaAs(002) Bragg reflection as well as comparably small changes of
the superlattice peak intensities. To classify the reliability of
our measurements, the experiment was repeated but without optical excitation, 
as shown on the right side of Figure~\ref{fig:setup}. The
difference pattern $I_\textbf{ne}-I_\textbf{ne}$, now being a
measure for the systematical errors of the experiment, identifies
only the shift of the Bragg reflection as genuine effect of the
optical excitation, whereas the altered intensities of the
superlattice peaks are predominantly due to statistical
fluctuations. A typical peak shift observed for a QD sample in the excited and non-excited state is shown in Figure~\ref{fig:data} (top left).


The shift of the GaAs(002) Bragg reflection corresponds to an
increase of the lattice constant of the GaAs matrix along the
heterostructure growth direction. To elucidate the origin of this
change, we studied the thermal expansion of the QD sample and,
serving as a reference, of a (001) oriented bulk GaAs sample of
comparable dimensions. We estimated possible laser excitation induced lattice heating effects 
by monitoring the energy of GaAs-related emission lines, namely of the band gap as well as defect-related transitions. Within the experimental accuracy of about 1 meV we did not resolve a shift of these lines, limiting possible crystal temperature increase to 10 K over the whole applied excitation power range. 
For comparison, the thermal expansion of both samples was determined by measuring the change of lattice constant $a$ 
along the [001] crystal direction from the GaAs(002) and GaAs(004)
Bragg reflections in the temperature range between 90~K and 125~K around the nominal sample temperature. 
These measurements were performed with a laboratory diffractometer in
$\theta-\theta$ geometry, which was equipped with an x-ray tube
emitting Cu K$_\alpha$ radiation of 8.048~keV photon energy. The
change of the perpendicular lattice constant $\Delta a/a_\text{s}$
normalized by the 300~K lattice constant is shown in Figure~\ref{fig:data} (top right). We performed a linear regression to the data in order to evaluate differences in the thermal expansion of the two samples (see dashed line for QD [001] and solid line for GaAs [001]) and found, that the thermal lattice 
expansion is hardly modified by the inclusion of the QDs. The change in slope of the linear fits for GaAs and QD sample is found to be $2\cdot10^{-7}$K${^{-1}}$ and relates to a lattice constant difference between GaAs and QD sample in the order of $2\cdot10^{-6}$ for $\Delta T = 10$K.   

Next we investigated the lattice expansion of the optically excited
samples along the [001] and [100] crystal directions as function of
laser power density $P$, where we first focus on the 532 nm
illumination. For that purpose high-resolution XRD measurements of
the GaAs(002) and GaAs(200) Bragg reflections were analyzed, also at
$T$ = 100~K. The measurement of the GaAs(200) reflection was
performed under a grazing incidence angle of
\mbox{$\alpha_\text{i}=0.5^\circ$}. The relative change of lattice
constant, $\Delta a/a_\text{s}(P)$, was obtained from determining
the laser-induced shift of the corresponding Bragg reflection. The
results of these measurements are shown in
Figure~\ref{fig:data} (bottom). Note that the excitation power
densities are corrected for the reflectivities in the setup.

For the GaAs sample, identical linear dependencies of the relative
lattice expansion on laser excitation power density are found for
the [001] and [100] crystal directions. The solid line Figure~\ref{fig:data} (bottom) shows 
the corresponding fit to the data.
Hence, the laser irradiation of the GaAs reference causes an
isotropic lattice expansion, as for thermal heating.

In contrast, the QD sample is characterized by a strongly
anisotropic lattice expansion, very different from the expected
isotropy of a pure thermal effect: while the excitation-induced
expansion along the [100] direction (dshed-dotted line) is similar to that in the GaAs 
sample, it is considerably enhanced along the [001] direction (dashed line). The
crystal therefore undergoes a tetragonal lattice distortion.

The data indicate that the in-plane effect occurs mostly
from thermal lattice expansion. On the other hand, the enhanced
expansion along the [001] direction apparently cannot be assigned to a sole
thermal effect. This is further corroborated by power dependent
measurements performed with the red laser emitting at 808~nm
wavelength, corresponding to a photon energy of $E=1.53$~eV. The
mechanism of lattice heating is quite different then, because the
carriers are excited only about one optical phonon energy (36.7~meV
in GaAs) above the GaAs band gap ($E_{\text{g,GaAs}}=1.50$~eV at
$T=100$~K), while for the green laser with $E=2.33$~eV photon energy
the excess energy of 0.8~eV corresponds to more than 20 optical
phonons. Once these energies have been released, carriers are
trapped in the wetting layer or highly excited QD states. From there,
further phonon emission has to occur to bridge the 0.4~eV energy
required for relaxation into the dot ground states, where the
photoluminescence shown in Fig.~1 is generated. For red light
excitation, the crystal heating therefore arises mostly from the QDs,
while for green illumination the heating occurs more homogeneously
in the crystal. Still we find for red light excitation that
the lattice expansion along the [001] direction is significantly
larger in the QD sample than for the GaAs reference. The independence of
the tetragonal QD lattice distortion on laser excitation wavelength
supports that it has an origin different from heating.
\cite{footnote}

We therefore suggest that the
anisotropic contribution to the lattice distortion is mostly induced by the
optically injected carriers after relaxation into their ground
states. This relaxation occurs on a timescale of a few ten ps. After
about 0.5 ns electrons and holes recombine radiatively, giving rise
to the photoluminescence in Fig.~1. However, the continuous wave
excitation maintains an on average steady carrier population in the
quantum dot ensemble. The electron-hole pairs couple strongly to the
lattice, leading to polaron formation.

This suggestion has to be tested regarding compatibility with the
experimental findings. Most importantly, there is the anisotropy of
the tetragonal lattice distortion. This anisotropy is in good accordance
with the expectation from the distribution of the electron and hole
wave functions in the QDs. From former studies, it is established
that there is a mismatch of the corresponding charge distributions
with the electron wave function located below that of the hole along
the vertical growth direction of (In,Ga)As self-assembled quantum
dots \cite{fry}. This implies that electron-hole pairs resemble
electric dipoles oriented along the [001] crystal direction. The
dipole orientation obviously facilitates and amplifies the polaron
formation along this direction.

Next, there is the excitation power dependence. The QD population by
electron-hole pairs in the ensemble is stochastic so that we monitor
an average distortion of the lattice by the polaronic effects. With
increasing excitation power, more QDs become populated by carriers
and contribute to the lattice distortion along the [001] direction
leading to an expansion increasing linearly with excitation
power, similar to the increase of photoluminescence emission
intensity. As function of excitation power it can be described by
\begin{align*}
    \Delta a/a_\text{s}(P) = (0.5\pm0.12)\cdot10^{-5} \cdot P,
\end{align*}
where the laser excitation power $P$ is measured in Wcm$^{-2}$. From
this average distortion one may estimate the optically induced
deformation at the QD layers. For simplicity, we assume this
distortion to be homogeneous in the different layers. An important boundary
condition is that we do see a shift of the X-ray reflections but no
significant change in peak intensity and width: this limits the
local strain differences to about $10^{-4}$. Together with the
average lattice expansion on the order of 2~fm per W/cm$^2$ and the
exponential x-ray penetration profile into the sample with an
absorption depth in the $\mu$m-range, we estimate an overall lattice distortion of
5~fm per W/cm$^2$ along the [001] direction at the QD layers. About
60\% of this distortion arise from polaronic effects.

In summary, we found clear evidence for an optically induced
tetragonal distortion of the GaAs crystal lattice into which
(In,Ga)As QDs are embedded. The anisotropy of the distortion arises
from polaronic effects initiated by QD confined carriers after their
optical excitation. We believe that these studies are
proof-of-principle studies and mark the beginning of x-ray studies
of 'condensed matter systems in operation'. So far, condensed matter
has been valuably studied by x-rays in passive mode, i.e. without
excitation, to understand their structure on an atomistic scale. The
impact of device operation on the lattice, especially under
extreme excitation conditions, opens a new dimension of their
understanding, and in particular, of the conditions that ultimately
lead to their failure. Many further developments in the analysis of
active devices can be foreseen such as higher spatial resolution
along the vertical direction, angular dependence, or temporal resolution where, for
example, the polaron formation dynamics in QDs is monitored.

We acknowledge the machine groups of HASYLAB and DELTA for providing
synchrotron radiation. We also acknowledge support by the BMBF
through grant no. 05K12PE1. S.\,T.\ and F.\,J.\ W.\ thank the NRW
Forschungsschule 'Forschung mit Synchrotronstrahlung in den Nano-
und Biowissenschaften' for financial support. J.\,M.\ (grant no. 05K10PEC) and A.\,D.\,W.\ (Wieck Q.com-H grant no. 16KIS0109) acknowledge financial support from BMBF. A.\,D.\,W.\ and M.\,T.\ thank the MERCUR foundation under contract number Pr-2013-0001.


\begin{thebibliography}{99}
\bibitem{footnote} A similar power level corresponds to a larger photon flux for the red illumination, while on the other hand the absorption
coefficient is smaller than for the green laser. The net effect of these factors is a similar number of electron-hole pairs in the QDs
for both excitation wavelengths, as evidenced from the photoluminescence spectra with comparable intensity in Fig.~1. Therefore, we expect 
comparable polaronic effects. Also the heating is likely comparable as both laser beams are completely absorbed in the crystals.
\end{thebibliography}
\end{document}